\documentclass[letter,10pt]{article} 
\usepackage{lscape,a4wide,amssymb,amsmath,cite,epsfig}
\newcommand{\be}{\begin{equation}} 
\newcommand{\ee}{\end{equation}} 
\newcommand{\bea}{\begin{eqnarray}} 
\newcommand{\eea}{\end{eqnarray}} 
\newcommand{\bean}{\begin{eqnarray*}} 
\newcommand{\eean}{\end{eqnarray*}}

\newcommand{\gapproxeq}{\lower 
.7ex\hbox{$\;\stackrel{\textstyle >}{\sim}\;$}} 
\newcommand{\lapproxeq}{\lower 
.7ex\hbox{$\;\stackrel{\textstyle <}{\sim}\;$}}

\newcommand{\bc}{\begin{center}} 
\newcommand{\ec}{\end{center}} 
\newcommand{\btab}{\begin{tabular}} 
\newcommand{\etab}{\end{tabular}}

\def\10bar{$\bar {\hbox{\bf 10}}$}
\def\3bar{$\bar {\hbox{\bf 3}}$}

\begin{document} 
 
\begin{titlepage} 
 
  \baselineskip=18pt \vskip 0.9in
\begin{center} 
  {\bf \Large Pentaquark Symmetries, Selection Rules and another potentially Narrow State}\\
  \vspace*{0.3in} {\large F.E. Close}\footnote{\tt{e-mail:
      f.close@physics.ox.ac.uk}}
  {\large and J.J.Dudek}\footnote{\tt{e-mail: dudek@thphys.ox.ac.uk}}\\
  \vspace{.1in} {\it Department of Physics - Theoretical Physics, University of Oxford,\\
    1 Keble Rd., Oxford OX1 3NP, UK} \\
  \vspace{0.1in}
 
\end{center} 
 
\begin{abstract} 
  
  We identify essential differences between the pentaquark and chiral
  soliton models of \10bar$_5$ and {\bf 8}$_5$ pentaquarks and
  conventional {\bf 8}$_3$ states, which are experimentally
  measurable. We show how the decays of $\Xi_5$ states in particular
  can test models of the pentaquarks, recommend study of the relative
  branching ratios of e.g. $\Xi^{-}_5 \to \Xi^-\pi^0:\Xi^0\pi^-$, and
  predict that the decay amplitude $\Xi_5 \to \Xi^*\pi$ is zero at
  leading order in pentaquark models for any mixture of \10bar and the
  associated {\bf 8}$_5$.  We also include a pedagogic discussion of
  wavefunctions in the pentaquark picture and show that pentaquark
  models have this {\bf 8}$_5$ with $F/D=1/3$, in leading order
  forbidding $\Xi_5 \to \Lambda K$.  The role of Fermi-Dirac symmetry
  in the $qqqq$ wavefunction and its implications for the width of
  pentaquarks are briefly discussed. The relative couplings 
  $g^2(\Theta_Q N K_Q^*)/g^2(\Theta_Q N K_Q) = 3$ for $Q \equiv s,c,b$.
  A further potentially narrow
  state $\Lambda$ in {\bf 8}$_5$ with $J^P = 3/2^+$ is predicted
  around 1650 MeV.
 
\end{abstract} 
\end{titlepage}

\subsection*{Introduction} 

The possible discovery of an exotic and metastable baryon with
positive strangeness, the $\Theta^+(1540)$, has led to an explosion of
interest in chiral soliton models (a version of which is cited as
having predicted this state) and their relation to quark models. In
this letter we propose explicit experimental tests that are sensitive
to the assumed dynamics and thereby can distinguish among models.
 
Such a state was predicted in a version of the chiral soliton model
\cite{dpp,skyrme} to be in a \10bar of flavour $SU(3)$ and to have
$J^P = 1/2^+$. Subsequent to the original observation\cite{leps},
several interpretations of the state have been suggested within quark
models (for an early review see \cite{fec03}). Their common feature is
that its constitution be $udud\bar{s}$ with one unit of orbital
angular momentum in the wavefunction; they differ in the ways that the
interquark dynamics causes the $1/2^+$ state to be the lightest and to
have an anomalously narrow width.

If the $\Theta^+$ is an isoscalar, then a common feature of all models
is that it is a member of a \10bar, which contains further exotic
states, $\Xi^{+,--}$. There are three main differences among the
implications of models that can distinguish among various dynamics.

(i) The magnitude of the mass gap spanning the \10bar from $\Theta$ to
$\Xi$ is significantly smaller in pentaquark
models than in the original formulation of the chiral soliton
model\cite{dpp,fec03,jw1,kl1} though the latter is somewhat flexible as has
recently been noted \cite{EKP};

(ii) The first excited state of $\Theta$ is predicted\cite{cd03c} in
pentaquark models to be a $J^P = 3/2^+$ isoscalar, in a $J^P = 3/2^+$
\10bar, whereas there is no place for such a state in the present
formulation of chiral soliton models;

(iii) The hadron decays of non-exotic members of the \10bar, in
particular those of $\Xi^{0,-}$ are especially sensitive to the
interquark dynamics in pentaquark models. A specific example has been
discussed in \cite{jw2} but we shall show here that there is a more
extensive set of relations and selection rules that arise in
pentaquark models
and which can discriminate among various dynamic and mixing schemes.
In particular the relative strengths of decays $\Xi^- \to
\Xi^-\pi^0:\Xi^0\pi^-$ test \10bar - {\bf 8} mixing\cite{jw2}; $\Xi^-
\to \Lambda K^-:\Sigma^0 K^-$ have selection rules that test the decay
dynamics that have been hypothesised\cite{maltman,fec03,carlson} to
suppress the pentaquark widths; and $\Xi \to \Xi^* \pi$ is predicted
to vanish for both \10bar {\bf and} ${\bf 8}_5$ initial pentaquark
states in such dynamics. The electromagnetic mass splittings of the
$\Xi $ states also contain important information.

(iv) In pentaquark models the spin-orbit forces imply the existence of
a nearby ${\bf 8}_5$ $J^P =3/2^+$ multiplet containing a $\Lambda$
that should be narrow and unmixed barring isospin violating effects.

We make some brief comments on point (ii) and then develop our main
thesis, which focuses on points (iii) and (iv).

{\bf 1. A low-lying $J^P = 3/2^+$ \10bar multiplet }\cite{cd03c}:

An essential difference between the pentaquark and chiral soliton
(Skyrme) models appears to be in their implications for the first
excited state of the $\Theta$.  In $qqqq\bar{q}$ with positive parity
$1/2^+$ there is necessarily angular momentum present, which implies a
family of siblings but with $J^P =3/2^+$.  The spin-orbit forces among
the quarks and antiquark lead to a mass gap between any member of the
$J^P =1/2^+$ and its $J^P =3/2^+$ counterpart, which was calculated in
ref.\cite{cd03c} to be significantly less than $m_{\pi}$ and possibly
only $O(10-50)$MeV in the models of \cite{jw1,kl1}.  Similar remarks
hold for all the members of the \10bar, such as $\Xi^{+,--}$, and
their non-exotic analogues that can also occur in {\bf 8}$_5$, such as
$\Xi^{0,-}$.  Such a \10bar family of $J^P = 3/2^+$ states does {\bf
  not} occur in the present formulation of chiral soliton models, nor
can it if the Wess-Zumino constraint selects allowed
multiplets\cite{Pra}.

In the Skyrme model there are exotic states with $J^P = 3/2^+$ or higher
but these are in {\bf 27} and {\bf 35} multiplets of $SU(3)_F$.  Such
states are also expected in pentaquark models (e.g.  isotensor
resonance with states ranging from $uuuu\bar{s}$ with charge +3 to
$dddd\bar{s}$ with charge -1)\cite{Gian}. The essential difference then is that
in the chiral soliton Skyrme models any spin 3/2 partner of the
$\Theta$ will exist in a variety of charge states with $I=1,2$ whereas
the unique feature of the pentaquark models\cite{jw1,kl1} is that the
first excited state is an {\bf isoscalar} analogue of the $\Theta$.
(There may be versions of pentaquark models where this state is higher
in mass but that it is isoscalar is universal in any quark model
description).

{\bf 2. Pentaquark wavefunctions, mixing and decays}:

In pentaquark models where the $(qqqq)$ is in $\bar{{\bf 6}}_F$, then
${\bar{\bf 6}} \otimes \bar{{\bf 3}} =$ \10bar $\oplus$ $ {\bf 8}_5$
leading to an {\bf 8}$_5$ that is degenerate with the \10bar$_5$
before mixing; chiral soliton models can accommodate an {\bf 8} (as a
radial excitation of the ground state nucleon octet) though
degeneracy is accidental. A challenge will be to decode the mixings
between \10bar$_5$, this {\bf 8}$_5$ and possible contamination with
excited {\bf 8}$_3$ in experiment.  This is our main focus.

The essential dynamics that underpins correlations among the flavours
and spins of quarks in QCD derives from a considerable literature that
recognises that $ud$ in colour $\bar{{\bf 3}}$ with net spin 0 feel a
strong attraction\cite{jaffe}. This might even cause the $S$-wave
combination to cluster as $[udu][d\bar{s}]$ which is the $S$-wave $KN$
system, while the $P$-wave positive parity exhibits a metastability such
as seen for the $\Theta$. Two particular ways of realising this are
due to Karliner and Lipkin\cite{kl1} and Jaffe and Wilczek\cite{jw1}.
 
In such models the basic correlation among quarks is to form
antisymmetric flavour pairs, in $\bar{\bf 3}$ of $SU(3)_F$.  In order
to study the decays and mixings of these states it is important to
have a well defined convention for their wavefunctions\cite{emails}.
We define the $\bar{{\bf 3}}_F = ({\bf 3}_F \otimes {\bf 3}_F)$ basis
states as
\bea
A \equiv (ud) \equiv (ud-du)/\sqrt{2} \sim \bar{s}  \nonumber \\
B \equiv (ds )\equiv (ds-sd)/\sqrt{2} \sim \bar{u}  \nonumber \\
C \equiv (su) \equiv (su-us)/\sqrt{2} \sim \bar{d} 
\eea 
for which
$U_-A = -C; V_-B = - A; I_-C = -B$. The $\Theta^+ \equiv AAA \equiv
(ud)(ud)\bar{s}$ and all other members of the \10bar follow by
operating on this state sequentially by $U_-$ and $I_-$ until all
states have been achieved. For reference they are listed in table 1.
We shall always understand the first two labels to refer to the
diquarks and the rightmost to refer to the antiquark in JW\cite{jw1},
and for KL\cite{kl1} the latter pair of labels is understood to be in
the triquark. The flavour correlations in the two models are thus
identical.

In addition to the three manifestly exotic combinations $AAA,BBB,CCC$
the nonexotic states can also form an octet. In the specific dynamics
advocated in\cite{jw1}, the quark pairs are strongly correlated into
scalar pairs with colour $\bar{{\bf 3}}$. These scalar ``diquarks" are
then forced to satisfy Bose symmetry, which leads naturally to the
following correlations. Their colour degree of freedom is
antisymmetric $\bar{{\bf 3}} \otimes \bar{{\bf 3}} \rightarrow {\bf
  3}$; their relative $L=1$ provides an antisymmetric spatial state;
their spin coupling is trivially symmetric; and Bose symmetry is
completed by their flavour pairings being symmetric. This leads
naturally to the positive parity \10bar. For the {\bf 8}$_5$ it leads
to the mixed symmetric ${\bf 8}^{M_S}$ states of table 1; in this
extreme dynamics there are no mixed antisymmetric ${\bf 8}^{M_A}$
analogues (e.g. $p^{} \equiv (AC-CA)A/\sqrt{2}$). This ${\bf 8}^{M_S}$
decays to {\bf 8}$\otimes {\bf 8} $ with $F/D=1/3$ as will become
apparent later. Similar occurs for the KL correlation where the
assumption that the triquark is in a ${\bf \bar{6}}_F$ implies that
the pentaquark system form \10bar $\oplus$ {\bf 8} with the same
symmetry type as in table 1.

Thus the selection rules that we obtain are common to all these
pentaquark models and a consequence of the assumed decay dynamics. The
proposal of refs.\cite{maltman,fec03,carlson} is that such pentaquarks
can naturally have narrow widths due to the mismatch between the
colour-flavour-spin state in an initial pentaquark and the
meson-baryon colour singlet states into which they decay.  For a
simple attractive square well potential of range 1fm the width of a
$P$-wave resonance 100MeV above $KN$ threshold is of order
200MeV\cite{jw1,maltman}. However, this has not yet taken into account
any price for recoupling colour and flavour-spin to overlap the
$(ud)(ud)\bar{s}$ onto colour singlets $uud$ and $d\bar{s}$ say for
the $KN$.
 
If decays are assumed to arise by
``fall-apart"\cite{maltman,fec03,carlson,buccella} without need for
gluon exchange to trigger the decay (even though gluon exchange may be
important in determining the eigenstates), then in amplitude, starting
with the Jaffe-Wilczek configuration, the colour recoupling costs
$\frac{1}{\sqrt{3}}$. It is further implicitly assumed that the
fall-apart decay to a specific channel occurs only when the
flavour-spin correlation in the initial wavefunction matches that of
the said channel. In such a case the flavour-spin correlation to any
particular channel (e.g. $K^+n$) costs a further
$\frac{1}{2\sqrt{2}}$, hence a total suppression in rate of
$\frac{1}{24}$. This was originally noted in \cite{maltman}.
 
The minimal assumption then is that a diquark must cleave such that
one quark enters the baryon and the other enters the meson. {\it While
  this is necessary, implicitly it is assumed also to be sufficient}:
any components in the wavefunction that are not kinematically allowed
to decay are assumed to be absolutely forbidden. Selection rules that
we obtain here assume this and therefore are implicitly a test of this
decay dynamics.  There is also a penalty for the spatial overlaps. If
once organised into colour singlets, the constituents then simply fall apart
in a $P$-wave with no momentum transfer, only the $L_z =0$ part of the wavefunction contributes.
This implies a further suppression from the $L=1 \otimes S=1/2 \to J
=1/2;3/2$ coupling. Thus a total suppression of 1/72 for the $1/2^+$
and 1/36 for $3/2^+$ may be expected\cite{cd03c}.
 
The general conclusion is that if such dynamics govern the decays,
then in such models a width of $O(1-10)$MeV for $\Theta \to KN$ may
be reasonable. The above dynamics also implies that 
$g^2(\Theta N K^*)/g^2(\Theta N K) = 3$ (this is also implicit in \cite{carlson}).
 Although the $NK^*$ decay mode is kinematically inaccessible this relation may
eventually be tested in photoproduction experiments\cite{qiang}. Analogously this
implies that $g^2(\Theta_c N D^*)/g^2(\Theta_c N D) = 3$. The $\Theta_c$ is predicted
in ref\cite{jw1} to lie below strong decay threshold but spin-orbit effects\cite{cd03c}
could elevate its mass such that it is even above $D^* N$ threshold (see e.g. \cite{kl1}).
Thus if $m(\Theta_c) > 2.95$ GeV, an enhanced intrinsic coupling to $D^* N$ could be
searched for.
 
With the wavefunctions in table 1 we can immediately account for the
relative strengths of final states by carefully exploiting the
symmetries of the wavefunctions. For example, the $\Theta \equiv
(ud-du)(ud-du)\bar{s}/2 \to [(ud-du)u][d\bar{s}]/2 -
[(ud-du)d][u\bar{s}]/2$ which maps onto $\Theta \to pK^0/\sqrt{2} -
nK^+/\sqrt{2}$.
 
These amplitudes for decays into meson ($M$) and baryon ($B$) also
depend on the flavour-spin symmetry of the baryon. If we make this
explicit ($\phi,\chi$ referring to the flavour and spin wavefunctions
respectively and $M_A,M_S$ denoting their mixed symmetry properties
under interchange\cite{fecbook}) we have
\[
({\bf \bar{3}_F},S=0)({\bf \bar{3}_F},S=0) \to
M+B(\phi^{M_A}\chi^{M_A}).
\]
The same colour-orbital  configuration for tetraquarks ($qqqq$) in overall spin $S=0$ can be
realised with diquarks in ${\bf 6_F},S=1$. The pattern of decays from
this configuration mirror those above except that the baryon's
flavour-spin symmetry is swapped
\[
({\bf 6}_F,S=1)({\bf 6}_F,S=1) \to M+B(\phi^{M_S}\chi^{M_S}).
\]
Thus if one imposed overall antisymmetry on the tetraquark
wavefunction one encounters for the flavour-spin part of the
wavefunction
\[
  |({\bf \bar{3}}_F,S=0)({\bf \bar{3}}_F,S=0) \rangle \pm |({\bf
    6}_F,S=1)({\bf 6}_F,S=1) \rangle .
  \]
Noting that there is an $L=1$ within the $(qqqq)$ system, the above
  wavefunctions imply that the (+) phase decays to $M + B({\bf 56})$
  in a $P$-wave and the (-) phase decays to $M + B({\bf 70}(L=1))$ in
  an $S$-wave.  The latter would naively be kinematically forbidden
  and as such lead to a suppressed width if the $\Theta$ were in this
  representation (which is in the {\bf 105} dimensional mixed symmetry
  representation of flavour-spin \cite{buccella}). However, one needs
  also to confront the kinematically allowed decays to $M + B({\bf
    56})$ from the (+) phase state (in the symmetric {\bf 126}
  representation and discussed in \cite{carlson}).  In practice
   decays shared by the $({\bf \bar{3}},S=0)({\bf \bar{3}},S=0)$
  and $({\bf 6},S=1)({\bf 6},S=1) $ states lead to mixing. If this is
  stronger than the mass gap between these two states, then one would
  obtain the above two configurations, leading to the possibility of 
  the $\Theta$ as a narrow
  state in {\bf 105} partnered by a (yet unobserved) broad partner in {\bf 126}
  (see also \cite{kl3}). By contrast, if
  the mixing is small on the scale of the mass gap, the wavefunction
  of the light eigenstate in this spin-zero tetraquark sector will be dominated by the $({\bf
    \bar{3}},S=0)({\bf \bar{3}},S=0)$ configuration, which is the
  Jaffe-Wilczek model\cite{jw1}. As noted in ref.\cite{maltman}, mixing with the
  spin-one tetraquark sector as manifested in the KL correlation can lead to a lower
  eigenstate. The discussion in the rest of our present paper does not depend on this dynamical question.
 
  Within the assumption that decays are driven by the fall-apart
  dynamics, the flavour patterns follow for all of these
  configurations.

  It is especially instructive to apply our study of the fall-apart to
  the $\Xi_5$ states. In what follows we assume that only the ground
  state baryon + $0^-$ meson channels are kinematically accessible. If
  other channels such as $1^-$ mesons could be accessed these would
  cause the intrinsic suppression to be less dramatic.
 
 \subsubsection*{2.1 Decays of $\Xi_5$ states}
 
 Starting with the wavefunction (see table 1)
\[
|\Xi^- ({\bf \bar{10}})\rangle =
-\frac{1}{2\sqrt{3}}\left([(ds-sd)(su-us)+(su-us)(ds-sd)]\bar{u}
  +(ds-sd)^2\bar{d} \right)
\]
we can rewrite this in flavour space in the form $(qqq)(q\bar{q})$.
\[
|\Xi^- ({\bf \bar{10}})\rangle = -\frac{1}{2\sqrt{3}}
\left([(ds-sd)s](u\bar{u} - d\bar{d}) + [(su-us)d +
  (sd-ds)u](s\bar{u}) - [(su-us)s](d\bar{u}) + [(ds-sd)d](s\bar{d})
\right)
\]
which maps onto the following ground state hadrons
\[
\Xi^- ({\bf \bar{10}}) \to -\frac{1}{\sqrt{6}}\left( \sqrt{2}
  \Xi^-\pi^0 +\Xi^0\pi^- - \sqrt{2}K^-\Sigma^0 -
  \bar{K}^0\Sigma^-\right)
\]
These agree in relative magnitudes and phases with the standard de
Swart results\cite{deswart,pdg}; they agree in relative magnitudes
with Oh et al.\cite{oh} but their phases differ from ours.  Refs
\cite{oh,jw2} do not discuss the ${\bf 8}$ decays as these depend in
general on an undetermined $F/D$.  However with the pentaquark
wavefunctions, as specified as in table 1, the octet from $\bar{{\bf
    6}}_F \otimes \bar{{\bf 3}}_F$ that is orthogonal to the \10bar is
\[
|\Xi_5^- ({\bf 8})\rangle =
-\frac{1}{2\sqrt{3}}\left([(ds-sd)(su-us)+(su-us)(ds-sd)]\bar{u} -
  \sqrt{2}(ds-sd)^2\bar{d} \right)
\]
and for the assumed decay dynamics employed in
\cite{maltman,fec03,carlson}, the particle decomposition is
\[
\Xi_5^- ({\bf 8}) \to -\frac{1}{\sqrt{24}}\left( \sqrt{6}\Xi^- \eta_1
  - \sqrt{3}\Xi^-\eta_8 - \Xi^-\pi^0 + \sqrt{2}\Xi^0\pi^- +
  2\sqrt{2}\Sigma^-\bar{K}^0 - 2 \Sigma^0K^- + 0\Lambda K^- \right)
\]
which corresponds to ${\bf 8} \to {\bf 8} \otimes {\bf 8}$ with
$F/D$=1/3 (or $g_1 = \sqrt{5}g_2$ in the de Swart
convention\cite{deswart,pdg}). With this one can therefore deduce the
branching ratios for $N,\Sigma,\Lambda$ states in ${\bf 8}_5$
immediately from existing tables\cite{deswart,pdg} and we do not
discuss them further here.

For the $\Xi_5$ we see immediately distinctions between the two
states.

(i) Isospin ($I=3/2$ versus $I=1/2$) is responsible for the distinctive
ratios
\[
\frac{\Gamma(\Xi_5^- \to \Xi^- \pi^0)}{\Gamma(\Xi_5^- \to \Xi^0 \pi^-)} = \left\{ \begin{array}{cc}
         1/2 & {\bf 8}\\
         2 & \bar{{\bf 10}} 
         \end{array} \right.
\]
and analogous for the $\Sigma K$ modes.

(ii) There is a selection rule that $\Lambda K^-$ modes vanish. For
the \10bar this is a trivial consequence of isospin; for the {\bf
  8}$_5$ it is a result of the pentaquark wavefunction, in particular
that the $qqqq$ flavour wavefunction of the pair of diquarks is
symmetric in flavour, (i.e. $\bar{{\bf 6}}_F = \bar{{\bf 3}}_F \otimes
\bar{{\bf 3}}_F$) leading to $F/D$=1/3.

A pedagogic explanation of the selection rule is as follows. The
$\Xi_5$ state wavefunctions contain two pieces of generic structure
$(dssu)\bar{u}$ and $(dssd)\bar{d}$. The $I=3/2$ and $I=1/2$ states
differ in the relative proportions of these two. However, only the
first component $(dssu)\bar{u}$ contains the $\bar{u}$ required for
the $K^-$ and this is common to both the $\Xi(I=3/2)$ and
$\Xi(I=1/2)$. Thus as the $\Xi(I=3/2) \to K\Lambda$ is trivially
forbidden by isospin, the $\Xi(I=1/2) \to K\Lambda$ must be also
unless there is cross-talk between the two components in the
wavefunction. This would happen if annihilation $(dssu)\bar{u} \to
(dss) \to (dssd)\bar{d}$ occurs. Thus observation of $\Lambda K^-$
could arise if there are admixtures of {\bf 8}$_3$ in the
wavefunction.

Rescattering from kinematically forbidden channels, such as $\Xi\eta$
can feed both $K\Sigma$ and $K\Lambda$, though this is not expected to
be a large effect if experience with light hadrons is relevant (such
as the small width of the $f_1(1285)$ not being affected by
rescattering from the kinematically closed $KK^*$ channel, and the
predicted $\pi_2 \to b_1 \pi \sim 0$\cite{barnes03} not being affected
by rescattering from the allowed channels $\pi f_2; \pi \rho $).
Whether this carries over to pentaquarks may be tested qualitatively
in models by comparing the relative suppression of $\Theta$,
$\Xi^{--}$ and $\Xi^-$ states; if there is no rescattering and the
$\Xi\eta$ channels are closed in the initial pentaquark wavefunction,
its width will be further suppressed from $1/24 \to \sim 3/115$ and
$K\Lambda \sim 0$. In this case the width of $\Xi^{-}$ (after phase
space effects have been removed) will be less than that of $\Xi^{--}$.
A dominance of $K\Lambda >K\Sigma$ can arise if there are pentaquark
configurations 
having $F=D$.  In this latter case the $\Sigma^0 K^-$ would be
forbidden but $\Lambda K^-$ allowed. The $\Lambda K:\Sigma K$ ratio in
general can be used to constrain the $F/D$ ratio and begin to
discriminate between various dynamical schemes.

(iii) Decays to $\Xi^*\pi$ and $\Sigma^* K$ for $\Sigma^*,\Xi^*$ in
the {\bf 10} are forbidden (even if allowed by phase space). For the
\10bar this is a result of \10bar $\neq$ {\bf 8} $\otimes$ {\bf 10} as
noted in ref\cite{jw2} who also discuss $SU(3)_F$ breaking as a
potential source of violation of this zero. However, this selection
rule may be stronger in the pentaquark models of refs.\cite{jw1,kl1}
due to the diquarks having antisymmetric flavour ($\bf \bar{3}$) and
spin zero, both of which prevent simple overlap of flavour-spin with
the ${\bf 10},S=3/2$ baryon decuplet resonances.  Thus although
$SU(3)_F$ allows ${\bf 8} \to {\bf 10} \otimes {\bf 8}$ to occur, for
the {\bf 8}$_5$ states of table 1 it is again forbidden as a
consequence of the antisymmetric flavour content of the wavefunction,
at least within the models of suppressed decay widths considered here.
While we discussed this for the Jaffe Wilczek wavefunction, Karliner
and Lipkin have one of their quark pairs strongly correlated into a
vector spin state within a triquark (e.g. $ud\bar{s}$) so the flavour
antisymmetries and explicit scalar diquark in the residual
wavefunctions suggest that this dynamics also would be challenged to
accommodate a violation of this selection rule.

The $I=3/2$ states will all be narrow. They are degenerate up to
electromagnetic mass shifts. Across the $I=3/2$ multiplet the mass
split is $\Xi^{--} - \Xi^+ = (d-u) + \langle e^2/R
\rangle$\cite{fecbook} where the Coulomb contribution in known hadrons
is $\sim 2-9$MeV, hence a spread of 3-10MeV is expected. For the non
exotic states $m(\Xi_{{\bf 8}_5}) > m(\Xi_{\bf \bar{10}})$, with
\[
m(\Xi^0_{{\bf 8}_5})-m(\Xi^0_{\bf \bar{10}}) = \frac{1}{2}\left[ m(\Xi^-_{{\bf 8}_5})-m(\Xi^-_{\bf \bar{10}}) \right] = \frac{1}{3} [ m(d) -
m(u)] \sim 1.5 - 2.5 \mathrm{MeV}
\] 
and hence degenerate to within better than 5 MeV. If the coupling to
$\Xi^* \pi$ vanishes for the ${\bf 8}_5$ as well as the \10bar, then
mixing by the common $\Xi \pi$ decay channels will be destructive. If
the widths are truly narrow the mass eigenstates become $\Xi_u \equiv
(ds)(su)\bar{u}$ and $\Xi_d \equiv (ds)(ds)\bar{d}$ separated by $\sim
10$MeV. The heavier state $b.r.(\Xi_d \to \Xi^-\pi^0)= 2\times b.r.(\Xi_d
\to \Sigma^-\bar{K}^0) $ (apart from phase effects) but it does not
decay to either $\Xi^0\pi^-$ nor $\Sigma^0 K^-$.  In contrast the
lighter state $\Xi_u \to \Xi^0\pi^-:\Xi^-\pi^0 = 2$, as for the pure
${\bf 8}$ (but with opposite relative phase), while it does not decay
to $\Sigma^-\bar{K}^0$.

Violation of these relations would imply either mixing with excited
{\bf 8}$_3$ states, be due to pentaquark components in the
wavefunction beyond those above, or because the width suppression is
realised by some dynamics other than implicit in
refs.\cite{maltman,fec03,carlson}. In the former case one would expect
the ${\bf 8}_3$ components to decay without suppression and dominate
the systematics of the widths. In this case there will be narrow
$\Xi^{-,0}$ with $I$=3/2 partnering the exotic $\Xi^{+,--}$ and broad
$I$=1/2 states that are akin to normal excited $\Xi$ states. By
contrast, were the $\Xi\pi$ charge ratios to show mixing between the
two $\Xi_5$ states with two narrow states such as $\Xi_d$ and $\Xi_u$,
then observation of any $\Lambda K$ or $\Xi^* \pi$ would require
components in the pentaquark wavefunction with different symmetries to
those above.

  \subsubsection*{2.2 Decays of $p_5$ and $n_5$ states}
  
  These follow immediately from $SU(3)$ tables with $F/D$=1/3.  In
  general there will be mixing between these as suggested by Jaffe and
  Wilczek. For the extreme $p_5(s\bar{s})$ and $p_5(d\bar{d})$ we have
\[
p_5(s\bar{s}) \to \frac{1}{2}\left(\frac{1}{\sqrt{2}}\Sigma^0K^+ +
  \sqrt\frac{3}{2} \Lambda K^+ - \Sigma^+K^0 + p \eta_{s} \right)
\]
where $\eta_s \equiv \eta_1/\sqrt{3} + 2\eta_8/\sqrt{6}$; and while
phase space only admits trivial $p_5(d\bar{d})$ decays to $N\pi$.

It is immediately apparent that the decays of $P_{11}(1440;1710)$ do
not fit well with this scheme.  First, there is a dominance of
nonstrange hadrons in the heavier $P_{11}(1710)$ with prominent
$\Delta \pi$ in the decays of both $P_{11}(1440;1710)$. This mode is
not possible for the $p_5$ states in \10bar nor in {\bf 8}$_5$ unless
overwritten by rescattering or mixing with {\bf 8}$_3$.

It is clear that $P_{11}(1440)$ is partnered by $P_{33}(1660)$ as in a
traditional {\bf 56} multiplet of SU(6) $qqq$ states. There is no
obvious sign of pentaquarks here.  A possibility is that the states
are linear combinations of $p_3 $ and $p_5$; the $p_5$ could even
dominate the wavefunction but its $O(1\mathrm{MeV})$ width is swamped by the
$O(100\mathrm{MeV})$ width of the unsuppressed $p_3$ component.  The $p_5$
decays listed above would then show up as rare decays at the $O(1\%)$
level.

\subsubsection*{3. $\Lambda_5$ state with $J^P =3/2^+$}

There is one further potentially narrow state in pentaquark models,
which has little opportunity for mixing with $qqq$ states. This is the
$\Lambda_5$ state that is the $J^P =3/2^+$ spin-orbit partner of
$\Lambda_5$ in ${\bf 8}_5$.

First note that \10bar contains $\Sigma_5$ but has no $\Lambda_5$.
The ${\bf 8}_5$ contains a $\Lambda_5$, and there will be no mixing
with \10bar so long as isospin is good.  If there were no mixing with
$\Lambda(qqq)$ excited states, this $\Lambda_5$ would be narrow, with
width identical to that of the $\Theta$ apart from phase space
factors.

The $\Lambda_5$ wavefunction shows that it has only one strange mass
quark and hence is similar to the $\Theta$ in this regard.
Ref.\cite{jw1} estimate $\sim 1600$ MeV for such a state (the excess
$\sim 60$ MeV relative to the $\Theta$ arising because the mass of a
$(us)\bar{d}$ set is larger than $(ud) \bar{s}$ due to the relatively
smaller downward mass shift in the $(us)$ diquark).  Scaling the
spin-orbit splitting from \cite{cd03c} and allowing for the relative
masses of the $\bar{s}/\bar{d}$ and $m(us)/m(ud)$ gives 40-70 MeV for
the $\Lambda_5$ mass gap of $3/2^+ -1/2^+$ and hence 1600-1700 as a
conservative estimate for the mass range for the partner $\Lambda_5
(3/2^+)$.

Perusal of the data\cite{pdg} shows that, for the $1/2^+$, mixing with
$qqq$ states is likely (given the existence of a candidate ${\bf
  56},0^+$ multiplet containing
$P_{11}(1440),\Lambda(1600),\Sigma(1660),\Xi(?)$).  However there is
no $3/2^+$ multiplet with a $\Lambda (1600-1700)$ seen, nor is one
expected in standard $qqq$ models. The first such is the set
containing $P_{13}(1720),\Lambda(1890) \cdots $. Thus 
there is a significant gap between $\Lambda(1890)$ and our predicted
$\Lambda_5(3/2^+)$.

The branching ratios for either the spin 1/2 or 3/2 states can be
determined from the breakdown
\[
\Lambda_5 \to -\frac{1}{2\sqrt{2}}\left(pK^- - n\bar{K}^0 +
  \Sigma^-\pi^+ - \Sigma^+\pi^- - \Sigma^0 \pi^0 -
  \sqrt{3}\Lambda\eta_{n\bar{n}} \right)
\]
Decays to $\Sigma^*\pi$ should be suppressed, even if they are
kinematically accessible.  The production rate of the spin 1/2 state
in $\gamma p \to K^+ \Lambda_5$ should be similar to that of $\gamma n
\to K^- \Theta$ (perhaps a factor of four smaller if K exchange
drives the production and $g(KN\Theta) = 2g(KN\Lambda_5)$). If the
arguments about $L \otimes S$ coupling and fall-apart dynamics are
correct, then we can expect the spin 3/2 state to be enhanced by a
factor of two relative to the spin 1/2 counterpart. A search in
$\gamma p \to K^+ \Lambda_5$ therefore seems appropriate.

If \10bar-${\bf 8}_5$ mixing is ideal, then also charged
$\Sigma_d^{\pm}$ states will occur which for $J^P=3/2^+$ should be
unmixed. For $J^P=1/2^+$, the amplitudes $g(\Theta^+K^+n) =
\sqrt{2}g(\Sigma_5^-K^-n)$ and so the relative photoproduction cross
sections should scale as $\sigma(\gamma n \to K^-\Theta^+) \sim
2\times\sigma(\gamma n \to K^+\Sigma_5^-)$\cite{qiang}. If the $\Sigma_5$ is mixed into
the $\Sigma(1660)$ then the latter state should be photoproduced at
least at the above rate and so may be a test for consistency.

\subsubsection*{Summary}

We advocate study of decays of the $\Xi$ states, especially the ratios of various charge modes, and
searching for $\Xi^* \pi$ and $\Lambda K$ as tests of the underlying dynamics that forms the states.
We also stress the importance of isolating the $J^P =3/2^+$ states that must occur in
\10bar and ${\bf 8}_5$ in pentaquark models but which have no analogue in the chiral soliton model.
These states are predicted to be within a few tens of MeV of their $1/2^+$ counterparts in
highly correlated models such as those of Jaffe-Wilczek or Karliner-Lipkin. Were the mass gap to be
significantly larger, then it could point to the presence of other components in the pentaquark 
wavefunction. By contrast, the
absence of such states together with the appearance of $J^P = 3/2^+$ in higher representations
such as {\bf 27} or {\bf 35} would support the chiral soliton models. 
 The $\Lambda_5(3/2^+)$ state may be relatively light and narrow and should be produced with
 similar strength to $\Theta$ in photoproduction. Its confirmation could play a significant role in helping to decode
 the mixing between pentaquarks and conventional states.


\begin{table}[!h]
\hspace{-1.5cm} \begin{tabular}{l|c|c}
  & \10bar & ${\bf 8}_5$ \\
\hline
$\Theta^+$ & $AAA$ &\\
\hline
$p$ & $-(ACA + CAA +AAC)/\sqrt{3}$ & $-(ACA +CAA -2AAC)/\sqrt{6}$\\
$n$ & $(ABA +BAA +AAB)/\sqrt{3}$ & $(ABA + BAA -2AAB)/\sqrt{6}$\\
\hline
$\Sigma^+$ & $(CAC+ACC+CCA)/\sqrt{3}$ & $(CAC+ACC -2CCA)/\sqrt{6}$\\
$\Sigma^0$ & $-(ABC+BAC+ACB+CAB+BCA+CBA)/\sqrt{6}$ & $-(ABC+BAC +ACB+CAB-2BCA-2CBA)/\sqrt{12}$\\
$\Lambda^0$ & & $-(ABC-ACB+BAC-CAB)/2$\\
$\Sigma^-$ & $(BAB+ABB+BBA)/\sqrt{3}$ & $(BAB+ABB-2BBA)/\sqrt{6}$\\
\hline
$\Xi^+$ & $-CCC$ &\\
$\Xi^0$ & $(CBC+BCC+CCB)/\sqrt{3}$ & $(CBC+BCC-2CBB)/\sqrt{6}$\\
$\Xi^-$ & $-(CBB+BCB+BBC)/\sqrt{3}$ & $-(CBB+BCB -2 BBC)/\sqrt{6}$\\
$\Xi^{--}$ & $BBB$ &
\end{tabular}
\caption{Pentaquark wavefunctions where $ABC$ are defined in the text. Note that consistency requires the
meson octet to be defined with each $q\bar{q}$ positive except for $\pi^+ = -u\bar{d}$; $\bar{K}^0 = - s\bar{d}$
and then $\pi^0 = (u\bar{u} - d \bar{d})/\sqrt{2}$. In this convention $\eta_8 = (2s\bar{s} - u\bar{u} - d\bar{d})/\sqrt{6}$
 For JW\cite{jw1} $A_1,A_2$ refer to diquarks and $A_3$ to the antiquark; for KL\cite{kl1} $A_1$ is diquark and
 $[A_2A_3]$ is the triquark in $\bar{{\bf 6}}_F$. The $\bar{\bf 6} \otimes \bar{\bf{3}}$ gives the \10bar and {\bf 8}
 as listed above.}
\end{table}

\bc 
{\bf Acknowledgements} 
\ec 
 This work is supported in part by grants from the Particle Physics and Astronomy
 Research Council and the EU-TMR program ``Euridice" HPRN-CT-2002-00311. We are 
 indebted to R.L.Jaffe and M.Karliner for comments on an early version of some of these ideas.



\begin{thebibliography}{99}

\bibitem{skyrme}
 A.Manohar, Nucl. Phys. B{\bf 248} 19 (1984);
 M.Chemtob, Nucl.Phys. B {\bf 256} 600 (1985); M.Karliner and M.Mattis,
 Phys. Rev D {\bf 34}, 1991 (1986); M.Praszalowicz in Proc. Cracow Workshop
 on Skyrmions and Anomalies, Mogilany, Poland, Feb 20-24, 1987, World Scientific
 1987, p.112
 \bibitem{dpp} D.Diakanov, V.Petrov and M.Polyakov, hep-ph/9703373; Z.Phys. A {\bf 359}
 (1997) 305


 
 \bibitem{leps} T.Nakano et al (LEPS Collaboration), Phys.Rev.Lett. {\bf 91} 012002
 (2003) hep-ex/0301020
 V.V.Barmin et al (DIANA Collaboration) hep-ex/0304040,
 S.Stepanyan et al (CLAS Collaboration) hep-ex/0307018,
J.Barth et al. ELSA hep-ex/0307083,

 
 \bibitem{fec03}F.E.Close, Conference summary talk in Proceedings of Hadron03, hep-ph/0311087
  
 \bibitem{jw1} R.L.Jaffe and F.Wilczek, Phys.Rev.Lett. {\bf 91}, 232003 (2003),hep-ph/0307341
 \bibitem{kl1} M.Karliner and H.J.Lipkin, Phys.Lett. B {\bf 575}, 249 (2003),hep-ph/0307243;hep-ph/0307343
 \bibitem{EKP} J.Ellis, M.Karliner and M.Praszalowicz, hep-ph/0401127

 \bibitem{cd03c} F.E.Close and J.J.Dudek, hep-ph/0311258, Phys.Lett.B (2004) in press.
  \bibitem{jw2} R.L.Jaffe and F.Wilczek, hep-ph/0312369
 \bibitem{maltman} B.Jennings and K.Maltman, hep-ph/0308286

 \bibitem{carlson} C.E.Carlson et al hep-ph/0312325

\bibitem{Pra} M.Praszalowicz, Phys.Lett.B {\bf 575}, 234 (2003)
\bibitem{Gian} R.Bijker, M.M.Giannini and E.Santopinto, hep-ph/0310281        
 \bibitem{jaffe} R.L.Jaffe, Phys.Rev. D {\bf 15} 281 (1977); R.L.Jaffe and F.E.Low, Phys.Rev D {\bf 19} 2105 (1979); F.E.Close and N.Tornqvist, J.Phys.G. {\bf 28} R249 (2002); K.Rajagopal and F.Wilczek hep-ph/0011333        
        
         \bibitem{emails} Unpublished discussions with R.Jaffe, M.Karliner and H.J.Lipkin
        
\bibitem{buccella} F.Buccella and P.Sorba, hep-ph/0401083

\bibitem{qiang} F.E.Close and Q.Ziang (in preparation)

 \bibitem{fecbook} F.E.Close, Introduction to Quarks and Partons, (Academic Press 1979)

\bibitem{kl3} M.Karliner and H.J.Lipkin, hep-ph/0401072


\bibitem{deswart} J.J.de Swart;Rev.Mod.Phys. {\bf 35} 916 (1963) 
\bibitem{pdg} Particle Data Group, Phys.Rev. D{\bf 66}, 010001 (2002) 
\bibitem{oh} Y.Oh et al hep-ph/0310117

\bibitem{barnes03} T.Barnes, rapporteur talk at Hadron03


\end{thebibliography}
\end{document}